\documentclass[prd,10pt,twocolumn]{revtex4-1}
\usepackage{amsmath}    
\usepackage{amssymb}	
\usepackage{graphicx}   
\usepackage{natbib}

\newcommand{\AdS}{{\rm AdS}}

\newcommand{\calA}{{\mathcal{A}}}
\newcommand{\calB}{{\mathcal{B}}}
\newcommand{\calC}{{\mathcal{C}}}
\newcommand{\calN}{{\mathcal{N}}}
\newcommand{\calO}{{\mathcal{O}}}
\newcommand{\calP}{{\mathcal{P}}}
\newcommand{\calR}{{\mathcal{R}}}
\newcommand{\calS}{{\mathcal{S}}}

\newcommand{\rmR}{{\rm R}}

\renewcommand{\d}{\partial}
\newcommand{\phee}{\varphi}
\renewcommand{\det}{{\rm det}\,}
\DeclareMathOperator{\tint}{{\textstyle\int}}
\DeclareMathOperator{\toint}{{\textstyle\oint}}
\DeclareMathOperator{\Det}{Det}
\newcommand{\tr}{{\rm tr}\,}

\newcommand{\eepsilon}{\varepsilon}
\newcommand{\Pexp}{\calP\!\exp}

\newcommand{\dblavg}[1]{\langle\!\langle #1 \rangle\!\rangle}   

\begin{document}
\title{Scaling dimensions in AdS/QCD and the gluon field strength propagator}
\author{Joshua W.\ Powell}
\email{jwp14@phy.duke.edu}
\affiliation{Department of Physics, Duke University, Durham, North Carolina 27708, USA}
\date{\today}
\begin{abstract}
  We derive the scaling dimension of antisymmetric tensor operators in the boundary theory of the AdS/CFT correspondence using a functional integral representation of the boundary-to-boundary propagators of their dual fields in the bulk.  We then apply this technique to AdS/QCD in which the bulk metric is warped, resulting in non-constant scaling dimensions.  In particular, we compute the two-point correlation function of gluon field strength operators, for which it is prerequisite to know the flow of the anomalous scaling dimension under rescaling.  The results are in very good agreement with quenched lattice QCD data, thus confirming the functional form of the scaling dimension.
\end{abstract} 
\maketitle

%

In this paper, we treat two closely related problems.  First we extend the formula for the scaling dimension of boundary operators known from gauge/gravity dualities between scale invariant theories to a specific correspondence between theories without scale invariance.   More precisely, we work with the extension of AdS/CFT to AdS/QCD, which has a warped background that breaks conformality and give an anomalous dimension to the field strength operators.  Then as an application we compute the two-point function of the gluon field strength operator which we compare with results from quenched lattice QCD.  We find that the AdS/QCD result is in very good agreement with lattice calculation.

We begin by establishing our technique in the familiar setting of the AdS/CFT correspondence between $\calN = 4$ supersymmetric Yang-Mills theory in four dimensions and supergravity on the background $\AdS_5\times S^5$ \cite{Maldacena:1997re,Witten:1998qj,Gubser:1998bc}.  This duality has opened up a new venue for probing strongly coupled behavior of quantum field theories.  The appearance of other correspondences relating gauge theories without supersymmetry or conformal symmetry to gravity duals has given hope that there could be a gravitational theory dual to QCD which describes its low energy, nonperturbative regime.  Some approaches to find such a dual QCD begin, as in the AdS/CFT correspondence itself, with a string theory in a background configuration of branes \cite{Witten:1998zw,Brower:2000rp,Polchinski:2001tt,Sakai:2004cn,Sakai:2005yt}.  Others instead posit a gravitational background assumed to encapsulate the scaling behavior of QCD in an approach to low energy QCD known as AdS/QCD \cite{Erlich:2005qh,DaRold:2005zs,Karch:2006pv,Andreev:2006ct,Forkel:2008un,Branz:2010ub,Vega:2010ne,BoschiFilho:2002vd}.  Its validity is tested by comparison with experiments or other calculations such as lattice QCD. In this paper, it is the framework we shall adopt.  One shortcoming of this approach is that it does not address the dynamical origin of the backgrounds considered.  We overlook this issue and take the background as given.

One of the earliest results of the AdS/CFT correspondence was a computation of the scaling dimensions of boundary operators \cite{Gubser:1998bc,Witten:1998qj}.  This was achieved by computing the Green functions of bulk fields dual to the operators.  We repeat the process here but use a functional integral expression of the propagator because this technique proves to be the most useful later on in the warped background of AdS/QCD. We focus on $\rmR$-symmetry singlet operators in the boundary theory so we may ignore the $S^5$ dimensions in the bulk spacetime.  Furthermore, we generalize slightly to a $d$-dimensional boundary theory, so that the dual gravitational theory has $\AdS_{d+1}$ background geometry.  We use coordinates in which the background metric takes the form
\begin{equation}\label{eqn:AdSMetric}
  ds^2 = g_{mn}dx^m dx^n = \frac{R^2}{z^2}\,(dz^2 + \delta_{\mu\nu}\,dx^\mu dx^\nu)
\end{equation}
and introduce the vielbein $v_m{}^a = (R/z)\delta_m{}^a$.  Note that we have Euclideanized the metric.  Throughout the paper, Greek indices denote all directions except the radial direction $z$, and lower case indices from the early alphabet label directions in the orthonormal frame.

A $(d-p)$-form operator on the boundary is dual to a $p$-form field $\phee^{(p)}$ in the bulk with action
\begin{equation}\label{eqn:FieldAction}
  S = \frac{1}{2}\int\!\phee^{(p)}_\calA(x)(\square^{(p)}_x + m^2)^{\calA\calB}\phee^{(p)}_\calB(x)\sqrt{g(x)}\,d^{d+1}x\,,
\end{equation}
where $\square^{(p)}_x$ denotes the Hodge Laplacian on a $p$-form field.  We are interested in the Green function of the differential operator in Eq.~\eqref{eqn:FieldAction} on the background of Eq.~\eqref{eqn:AdSMetric}.  That is, we seek the solution to
\begin{equation}\label{eqn:BiformGreenFuncDefn}
  (\square^{(p)}_x + m^2)_{\calA}{}^{\calB} G^{(p)}_{\calB\calC}(x,y) = g(x)^{-1/2}\delta_{\calA\calC}\,\delta^{(d+1)}(x,y)\,,
\end{equation}
where $\calA$, $\calB$ and $\calC$ are indices appropriate for a $p$-form representation and $\delta^{(d+1)}(x,y)$ is the Dirac $\delta$ function on the curved space.  Weitzenb\"ock identities relate the Hodge Laplacian on $p$-forms to the componentwise application of the standard Laplace-Beltrami operator (scalar Laplacian) plus curvature terms \cite{Labbi:2006}.  Due to the maximal symmetry of the AdS background only the Ricci scalar curvature $\calR$ appears.  Its coefficient reflects the dimension of the space and the degree of the form field.  More specifically, Eq.~\eqref{eqn:BiformGreenFuncDefn} becomes
\begin{align}\label{eqn:ComponentDiffEqn}
  \Big(\square_x + m^2 &+ \frac{p(d+1-p)}{d(d+1)}\,\calR(x)\Big)\,G^{(p)}_{\calA\calB}(x,y)\\ &\hspace*{2cm} = g(x)^{-1/2}\delta_{\calA\calB}\,\delta^{(d+1)}(x,y)\,,\nonumber
\end{align}
where $\square_x$ is now the scalar Laplacian acting on the components of the Green function \cite{Folacci:1990ea,Bena:1999be,Naqvi:1999va}.  Absent background fields which designate special directions to reference, the Green function has a tensor decomposition which must respect the isometries of the Euclidean AdS background.  The general result is given in Refs.~\cite{Bena:1999be,Naqvi:1999va}, but in the limit that both $x$ and $y$ sit near the conformal boundary it reduces to
\begin{equation}\label{eqn:TensorDecomp}
G^{(p)}_{\calA\calB}(x,y) = \delta_{\calA\calB}\,G^{(p)}(x,y) + (\text{boundary term})\,,
\end{equation}
where the function $G^{(p)}(x,y)$ satisfies
\begin{align}\label{eqn:ComponentAction}
  \Big(\square_x + m^2 &+ \frac{p(d-p)}{d(d+1)}\,\calR(x)\Big)G^{(p)}(x,y)\\ &\hspace*{2cm} = g(x)^{-1/2}\delta^{(d+1)}(x,y)\,.\nonumber
\end{align}
The boundary terms in Eq.~\eqref{eqn:TensorDecomp} are total divergences and serve to adjust the boundary conditions of the Green function at infinity but will not affect our analysis.

In Refs.~\cite{Bena:1999be,Naqvi:1999va}, computation of $G^{(p)}(x,y)$ hinged on the ability to solve the relevant differential equations directly, which is not possible in the deformed background we will consider later.  Instead, a general technique to solve differential equations is to use functional integrals to invert the differential operator thereby providing a formal solution.  We will employ this approach and then make the functional integral tractable using a saddle point approximation, or equivalently the small noise, adiabatic or WKB approximation  \cite{Schwinger:1951nm, Bekenstein:1981xe, Stephens:1988jm}.  This approximation is exact in the $\AdS_{d+1}$ background because it is homogenous and isotropic.  The approximation will be good but not exact in the warped, asymptotically $\AdS_{d+1}$ background considered later.  Path integral solutions are most often seen in physics as solutions of the Schr\"odinger equation, but the technique is also applicable in a classical setting such as the bulk field theory considered in this paper.  We use a Feynman-Schwinger representation of the Green function, which expresses the dynamics of the one-particle sector of a field theory in terms of the particle's worldline \cite{Schwinger:1951nm,Bekenstein:1981xe,Bastianelli:2002fv}:
\begin{widetext}
  \begin{equation}\label{eqn:SchwingerRep}
    G^{(p)}(x,y) =\int\limits_{Z(0)=y}^{Z(1)=x}\![dZ(\lambda)\,de(\lambda)] \,\exp\bigg(\!-\frac{1}{2}\int_0^1\! \bigg[e^{-1}\dot Z^a \dot Z^b \delta_{ab}(Z) + e\,(m^2-\kappa_{d,p}\calR)\bigg]\,d\lambda\bigg)\,.
  \end{equation}
\end{widetext}
Here $Z(\lambda)$ is a parametrization of the path followed by the particle, $e$ is an einbein field transforming under reparametrization of the worldline as $e^\prime\,dZ^\prime = e\,dZ$, and $\calR$ is the Ricci scalar curvature of the background.

Whether it is used for a classical or quantum system, one must pick a discretization scheme of a functional integral to define it \footnote{This statement is analogous to the need to fix an ordering scheme when treating a problem using an operator formalism.  Even classical systems have this issue, which ultimately comes from techniques in functional analysis and is not unique to quantum dynamics.}.  In a curved background different schemes give rise to different couplings $\kappa_{d,p}$ to the curvature.  However the midpoint scheme uniquely possesses covariance under general coordinate transformation \cite{Grosche:1987ba,Bastianelli:2005vk}.  We demand coordinate invariance and therefore choose the midpoint scheme.  We add to the Euclideanized path integral action the curvature coupling it induces \cite{Grosche:1987dq,Grosche:1987de},
\begin{equation}\label{eqn:DiscretizationTerm}
  \Delta S = -\frac{1}{2}\int_0^1 \frac{(d/2)^2}{d(d+1)}\,\calR\,e\,d\lambda\,.
\end{equation}
The specific form of this extra term is appropriate for the saddle point approximation and moves the metric dependence from the path integral measure up into the action \cite{Gutzwiller:1985,Grosche:1987de}.  Equation~\eqref{eqn:DiscretizationTerm} combines with the curvature term from Eq.~\eqref{eqn:ComponentDiffEqn} to determine
\begin{equation}
  \kappa_{d,p} = -\frac{p(d-p)}{d(d+1)} + \frac{(d/2)^2}{d(d+1)} = \frac{1}{d(d+1)}\Big(\frac{d}{2}-p\Big)^2\,.
\end{equation}

To compute Eq.~\eqref{eqn:SchwingerRep} we make the small fluctuation approximation by expanding Eq.~\eqref{eqn:SchwingerRep} around the path $Z_0$ with minimum action $S_w^{(0)}$ and treating the fluctuations only to quadratic order \cite{Stephens:1988jm}.  To do this, first the particle's trajectory is decomposed as
\begin{equation}
  Z^a(\lambda) = Z^a_0(\lambda) + \zeta^a(\lambda)\,,
\end{equation}
imposing the orthogonality constraint $\delta_{ab} \dot Z_0^a(\lambda) \zeta^b(\lambda) = 0$ for all $\lambda$.  This means $\zeta(\lambda)$ has $d$ fluctuating degrees of freedom.  Next we rewrite Eq.~\eqref{eqn:SchwingerRep} in terms of the fluctuation $\zeta^a(\lambda)$ and truncate the action to quadratic order.  The zeroth order term is the minimum action
\begin{equation}\label{eqn:ClassicalAction}
  S_w^{(0)}(x,y|s) = \frac{\sigma(x,y)^2}{2s} + \frac{s}{2}(m^2 - \kappa_{d,p}\calR)\,,
\end{equation}
where $\sigma(x,y)$ is the proper distance of $Z_0(x,y)$ and $s = \int_0^1 e(\lambda)\,d\lambda$.  The term linear in $\zeta^a$ vanishes by construction. The quadratic term captures the effects of fluctuations around the geodesic.  This term must be kept as the path integral is dominated by paths with finite quadratic variation.  That is
\begin{align}
  &D_{\rm VM}(x,y|s)^{1/2}\nonumber \\
  &= \!\!\int\limits_{\zeta(0)=0}^{\zeta(1)=0}\!\![d\zeta(\lambda)]\exp\bigg(\frac{1}{2s}\int_0^1\!\zeta^a\,(\delta_{ab}\d_\lambda^2 + M_{ab})\zeta^b\,d\lambda\bigg)\label{eqn:FlucPathInt}\\
  &= \frac{1}{(2\pi s)^{(d+1)/2}}\bigg(\frac{\Det \det^\prime(-\delta_{ab}\d_\lambda^2 - M_{ab})}{\Det \det^\prime(-\delta_{ab}\d_\lambda^2)}\bigg)^{-1/2}\label{eqn:AdSVMD}
\end{align}
where the Riemann curvature tensor $\calR_{abcd}$ is used to define
\begin{equation}
  M_{ab} = \calR_{acbd}\dot Z_0^c \dot Z_0^d\,.
\end{equation}
Here $\Det$ is the functional determinant.  There is a zero mode of the worldline action stemming from reparametrization invariance.  It contributes to the prefactor in Eq.~\eqref{eqn:AdSVMD}.  But we must regulate the functional determinants to exclude it, which we do by defining $\det^\prime$ to be the determinant only over the subspace orthogonal to $Z_0^a$.  We recognize the action for the fluctuation field $\zeta^a$ as the geodesic deviation equation, also called the orthogonal Jacobi equation.  Higher orders of this equation correspond to the generalized Jacobi equation \cite{Colistete:2002ka,Perlick:2007ux}, and are proportional to derivatives of certain components of the curvature tensor, up to boundary terms in the action.  In the case of the unwarped AdS background the higher-order terms vanish identically.  In the warped case considered later, the higher-order terms vanish asymptotically for large and small separation of the end points and stay much smaller than the quadratic terms for intermediate distances.  For consistency of notation with much of the literature, we have written the fluctuation determinant in terms of $D_{\rm VM}(x,y|s)$, the Van Vleck-Morette determinant whose square root is the fluctuation determinant \cite{Bekenstein:1981xe,Stephens:1988jm,PhysRevD.28.417}.

Notice that both Eqs.~\eqref{eqn:ClassicalAction} and \eqref{eqn:AdSVMD} depend on $e(\lambda)$ only via the modulus $s = \int_0^1 e\,d\lambda$.  After fixing to the gauge $e(\lambda) = s$, the functional integral over the einbein reduces to a Riemann integral over $s$:
\begin{equation}\label{eqn:SemiclassicalGreenFunc}
  G^{(p)}(x,y) = \int_0^\infty\!ds\,D_{\rm VM}(x,y|s)^{1/2}\,\exp(-S_w^{(0)}(x,y|s))\,.
\end{equation}

The $\AdS_{d+1}$ metric of Eq.~\eqref{eqn:AdSMetric} has Riemann curvature tensor
\begin{equation}\label{eqn:AdSCFTRiemannTensor}
  \calR_{abcd} = -\frac{1}{R^2}\,(\delta_{ac}\delta_{bd} - \delta_{ad}\delta_{bc})\,.
\end{equation}
If $\lambda$ is an affine parameter then the tangent vector $\dot Z^a_0(\lambda)$ has constant length
\begin{equation}\label{eqn:AffineParam}
  \delta_{ab}\dot Z^a_0(\lambda) \dot Z^b_0(\lambda) = \sigma(x,y)^2\,.
\end{equation}
Using this fact and $\delta_{ab} \dot Z_0^a(\lambda) \zeta^b(\lambda) = 0$, we find
\begin{equation}\label{eqn:AdSTidalTensor}
  M_{ab} = -\delta_{ab}\,\sigma(x,y)^2/R^2\,.
\end{equation}
We substitute Eq.~\eqref{eqn:AdSTidalTensor} into Eq.~\eqref{eqn:AdSVMD} and use the fact that the differential operator is proportional to $\delta_{ab}$ to write
\begin{align}\label{eqn:DetOfAPower}
  \bigg(&\frac{\Det \det^\prime(-\delta_{ab}\d_\lambda^2 - \delta_{ab}\sigma(x,y)^2/R^2)}{\Det \det^\prime(-\delta_{ab}\d_\lambda^2)}\bigg)\\ &\hspace*{1.2cm}= \bigg(\frac{\Det(-\d_\lambda^2 - \sigma(x,y)^2/R^2)}{\Det(-\d_\lambda^2)}\bigg)^d\,.\nonumber
\end{align}
This step is the assertion that any sensible regularization of the functional determinant respects the degeneracy of eigenvalues resulting from the $d$ identical copies of the second-order differential operator induced by the $\det^\prime$ on the left-hand side.  A proof of this fact using zeta regularization is given in Ref.~\cite{Elizalde:1997nd}.  There are numerous ways to compute the ratio of the functional determinants of these second-order differential operators.  We use the Gel'fand-Yaglom theorem \cite{Gel'fand:1959nq,Dunne:2007rt} because this technique is useful later.  This theorem states that the normalized determinant of an operator $-\d_\lambda^2 - V(\lambda)$ acting on the space of functions supported on the interval $\lambda \in [0,1]$ with vanishing boundary values is
\begin{equation}
  \frac{\Det(-\d_\lambda^2 - V(\lambda))}{\Det(-\d_\lambda^2)} = u(1)^{-1}\,,
\end{equation}
where
\begin{equation}
  (\d_\lambda^2 + V(\lambda))u(\lambda) = 0
\end{equation}
with boundary conditions $u(0) = 0$ and $\dot u(0) = 1$.  For a constant $V = -\sigma(x,y)^2/R^2$ it follows immediately that
\begin{align}
  D_{\rm VM}&(x,y|s)^{1/2}\nonumber\\
    &= \frac{1}{(2\pi s)^{(d+1)/2}}\bigg(\frac{\sigma(x,y)/R}{\sinh(\sigma(x,y)/R)}\bigg)^{d/2}\,.\label{eqn:AdSVDMExplicit}
\end{align}

Using Eqs.~\eqref{eqn:ClassicalAction} and \eqref{eqn:AdSVMD} in Eq.~\eqref{eqn:SemiclassicalGreenFunc} shows that the saddle point approximation to the Green function takes the form
\begin{align}
  G^{(p)}(x,y) &= \calN K_{(d-1)/2}\bigg(\sigma(x,y)\sqrt{m^2 - \kappa_{d,p}\calR}\bigg)\label{eqn:Saddle Point}\\
  &\times\bigg(\frac{\sigma(x,y)/R}{\sinh(\sigma(x,y)/R)}\bigg)^{d/2}\bigg(\frac{\sqrt{m^2 - \kappa_{d,p}\calR}}{\sigma(x,y)} \bigg)^{(d-1)/2}\,.\nonumber
\end{align}
$\calN$ is a normalization factor and $K_{(d-1)/2}$ is a modified Bessel function.  Since we are computing the boundary-to-boundary propagator, put $x = (x^\mu,\eepsilon)$ and $y = (y^\mu,\eepsilon)$, for $\eepsilon$ positive but infinitesimal.  The geodesic distance is
\begin{equation}
  \label{eqn:BulkBndyDistFunc}
  \sigma(x,y) = R\,\ln\frac{|x^\mu - y^\mu|^2}{\eepsilon^2} + O(\eepsilon)\,.
\end{equation}
As $\eepsilon$ tends to 0, $\sigma(x,y)$ diverges and the $\sinh$ function and the Bessel function in Eq.~\eqref{eqn:Saddle Point} approach their asymptotic forms: $\sinh x \approx e^x/2$ and $K_{(d-1)/2}(x) \approx \sqrt{\pi/2x}\,e^{-x}$ for any $d$.  The asymptotic limits of the factors combine to eliminate dependence on $\sigma(x,y)$ everywhere except in the exponent.  We regularize $\sigma(x,y)$ at both end points with some mass scale $\mu$ by setting $\sigma(x,y) = \sigma_{\rm reg}(x,y) - 2 R \ln (\mu\eepsilon)$ and absorbing the divergent term into the normalization, $\calN \rightarrow \widetilde\calN(\mu)$.  The resulting Green function is
\begin{align}\label{eqn:AdSProp}
  G^{(p)}_{\rm reg}(x,y) &= \widetilde\calN\,\exp(-\Delta\,\sigma_{\rm reg}(x,y)/R)\\
  &= \frac{\widetilde\calN}{|x^\mu-y^\mu|^{2\Delta}}\,,\nonumber
\end{align}
with exponent
\begin{equation}\label{eqn:ScalingDimFormula}
  \Delta = \frac{d}{2} + \sqrt{(mR)^2 - \kappa_{d,p}R^2\calR}\,.
\end{equation}
The first term on the right comes from the fluctuation determinant and the second term comes from the minimum action.  After substituting $\calR = -d(d+1)/R^2$ and 
\begin{equation}
  \kappa_{d,\,p} = \frac{1}{d(d+1)}\,\bigg(\frac{d}{2} - p\bigg)^2\,,
\end{equation}
into Eq.~\eqref{eqn:ScalingDimFormula}, we recover the familiar scaling dimension formula of AdS/CFT \cite{Gubser:1998bc,Witten:1998qj}.

In Ref.~\cite{Minahan:2012fh} the Green function of a bulk field is also computed.  The author notes that the particle worldline representation of the propagator in AdS behaves as though $\Delta/R$ appears in place of the mass $m$ in the original field action.  Here, we have shown that a path integral computation of the two-point function can generate the needed terms to convert $m$ in the field Lagrangian into the correct expression for $\Delta/R$ in the worldline representation.  The path integral technique frees us from the need to solve the Green function's differential equation by inspection and allows us to work in more complicated backgrounds.

References~\cite{Papadimitriou:2004ap,Papadimitriou:2004rz} provide a treatment of general two-point functions in modified, asymptotically AdS backgrounds using a generalized Hamiltonian approach.  A prescription for generating renormalized bulk actions is presented.  Much of the analysis involves effects from a dynamical background geometry.  In this paper, the geometry is taken to be static, thereby greatly simplifying the situation.  Further simplifications in the present paper arise from restricting focus to the single particle sector of the bulk theory.

We now apply the path integral approach to compute the field strength two-point function in AdS/QCD.  The technique is largely the same but we must introduce two changes to compare our results with the lattice.  First we note that Yang-Mills field strength operators are not gauge invariant.  In order to have a gauge invariant operator, we introduce a Wilson loop with a gauge transformation which cancels that of the field strength operators.  Thus the first change we make is to add extended rather than only pointlike operator insertions in the boundary gauge theory.  The loop current on a closed contour $\calC = \{c(\tau)\,:\,\tau\in[0,1]\}$ is
\begin{equation}\label{eqn:LoopCurrent}
  j_\calC^\mu(x) = \int_0^1 \dot c^\mu(\tau)\,\delta^{(d)}(x - c(\tau))\,d\tau\,.
\end{equation}
The current enters the functional integral via the factor $\exp(-S_{\rm int}[A,j]) = \exp(-\int A \cdot j\,dx)$, the standard minimal coupling of a conserved current to a vector gauge field.  In non-Abelian theories, the exponential is understood to be path ordered.  We use a shorthand notation for the expectation value of a Wilson loop on $\calC$ dressed by inserting operator $\calO$:
\begin{equation}
  \dblavg{\calO}_\calC = N_c^{-1}\int[dA]\,e^{-S_{\rm YM}}\,\tr\calP\{\calO\,\exp(-\toint_\calC A(x) \cdot dx)\}\,.
\end{equation}
The trace is over gauge indices and $\calP$ denotes path ordering.  We note that
\begin{equation}
  Z[j_\calC] = \dblavg{1}_\calC\,;
\end{equation}
that is, the Wilson loop with no additional operator insertions is precisely the generating functional $Z[j_\calC]$.  Rewrite $\dblavg{1}_\calC$ using the non-Abelian Stokes' theorem:
\begin{align}
  \dblavg{1}_\calC &= \int[dA]\,e^{-S_{\rm YM}}\,\Pexp\big(-\toint_\calC A \cdot dx\big)\nonumber\\ &= \int[dA]\,e^{-S_{\rm YM}}\,\Pexp\big(\!-\toint_\calS F \cdot  d^2a \big)\,.\label{eqn:NonAbStokes}
\end{align}
Here $\calS$ is a surface with boundary $\calC$ and differential area 2-form $d^2 a^{\mu\nu}$.  In lattice QCD, the last expression of Eq.~\eqref{eqn:NonAbStokes} is discretized and computed numerically.  The functional integral over different configurations of the gauge field corresponds to a sum over states with different plaquettes contributing to the action.  The contributing plaquettes combine to form the surface $\calS$.  In the strong coupling limit, the minimal area surface dominates the integral \cite{Smit:2002ug}.  This picture will be connected with the dual theory.
 
The normalized two-point function of field strength operators connected by the Wilson loop along $\calC$ is
\begin{equation}\label{eqn:DFF}
  G^{(FF)}_{\mu\nu\rho\sigma}(x,y|\calC) = \dblavg{F_{\mu\nu}(x)F_{\rho\sigma}(y)}_\calC/\dblavg{1}_\calC\,.
\end{equation}
Gauge invariance requires that the points $x$ and $y$ must lie on $\calC$.   To compute $\dblavg{FF}_\calC$ in Eq.~\eqref{eqn:DFF} we note that the insertion of $F_{\mu\nu}$ at a point along $\calC$ can be achieved by taking the area derivative of the contour at this point \cite{Mandelstam:1968hz, Makeenko:1979pb}.  We modify the contour $\calC$ by adding an infinitesimal loop $\delta c^{\mu\nu}(x)$ at point $x$ which spans the directions $\mu$ and $\nu$.  Doing so induces a change in the loop current operator to $j_\calC + \delta j_\calC(x)$ and a corresponding shift in the partition function:
\begin{equation}
  \dblavg{1}_{\calC + \delta\calC} = Z[j_\calC + \delta j_\calC(x)] \approx \dblavg{1-\delta c^{\mu\nu}(x)F_{\mu\nu}(x)}_\calC\,.
\end{equation}
We extract the second term on the right using a loop or area derivative,
\begin{equation}
  \dblavg{F_{\mu\nu}(x)}_{\calC} = \frac{\delta Z[j_\calC]}{\delta c^{\mu\nu}(x)}\,.
\end{equation}
Two area derivatives insert two field strength operators:
\begin{equation}\label{eqn:TwoPtFuncInBackground}
  \dblavg{F_{\mu\nu}(x)F_{\rho\sigma}(y)}_\calC = \frac{\delta^2 Z[j_\calC]}{\delta c^{\mu\nu}(x)\delta c^{\rho\sigma}(y)}\,.
\end{equation}
On the lattice, an area derivative amounts to adding an extra plaquette to the contour $\calC$.  

We now compute $\dblavg{1}_\calC$ and $\dblavg{FF}_\calC$ using the standard treatment of Wilson loops in the AdS/CFT and then the AdS/QCD correspondence.  In the dual in the bulk theory of a Wilson loop on the boundary is a classical Nambu-Goto string ending on $\calC$ \cite{Maldacena:1998im}:
\begin{equation}
  \dblavg{1}_\calC = \min_{X_\calC} \exp(-S_{NG}[X_\calC])\,,
\end{equation}
where the Nambu-Goto action is given by
\begin{equation}\label{eqn:NGAction}
  S_{\rm NG}[X] = \frac{1}{2\pi\alpha^\prime}\int d^2\xi\,\sqrt{\smash{\underset{\alpha,\beta}{\det}}\,g_{mn}(X)\d_\alpha X^m \d_\beta X^n}
\end{equation}
and the minimum is over all worldsheets $X_\calC$ with boundary fixed to be $\calC$.  
\begin{figure}[t]
  \centerline{
    \includegraphics[width=9cm,trim=1cm 25cm 9.5cm 1cm,clip]{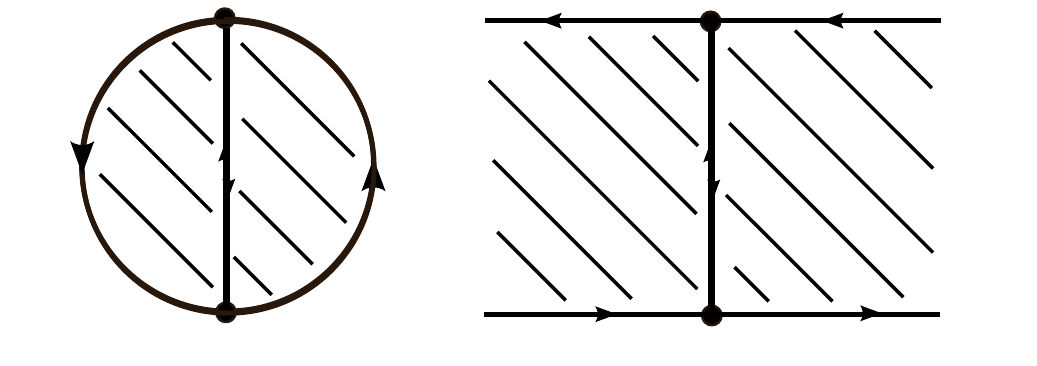}
  }
  \caption{Two possible contours $\calC$.  The contour on the right is the one used in our computations.}
  \label{fig:Contours}
\end{figure}

We return briefly to the gauge theory on the boundary.  The second area derivative of $Z[j_\calC]$ is the two-point function of the field strength in the gauge field background with source supported along the contour $\calC$.  The correlation of adjoint operators along the Wilson loop is realized in a lattice discretization of the gauge theory as coincident fundamental and antifundamental parallel transporters \cite{DiGiacomo:1992df,DiGiacomo:1996mn,Laine:1997nq}.  The antifundamental parallel transporter completes one half of the line current $j_\calC$ and the fundamental parallel transporter completes the other half, as shown in Fig.~\ref{fig:Contours}.  The propagator of the field strength operator should be thought of as a feature of the surface of the Wilson loop.  This picture describes the gauge invariant propagation of a gauge noninvariant quantity such as the field strength.  The corresponding picture in the bulk theory is that the propagator of the dual of the field strength operators may not propagate just anywhere in the bulk but must have a worldline lying within the worldsheet of the Nambu-Goto string dual to the Wilson loop.  That is, we look at the second-order response of the Nambu-Goto action to two wrinkles along its boundary inserted at the location of the field strength operators.  This is described as the propagation of the gauge noninvariant field within the Nambu-Goto worldsheet \cite{Drukker:1999zq}.

As an example of computing a boundary field strength correlator using AdS/CFT, Ref.~\cite{Polyakov:2000ti} computes the second area derivative of a straight, infinite Wilson line, with the result
\begin{equation}
  \dblavg{F_{\mu\nu}(x)F_{\rho\sigma}(y)}_\calC \propto 1/|x - y|^4\,,
\end{equation}
as expected.  Similarly, Ref.~\cite{Miwa:2006vd} treats the insertion along a circular Wilson loop of composite operators constructed from the scalar superpartner of the gauge field.  Both Refs.~\cite{Polyakov:2000ti} and \cite{Miwa:2006vd} present the picture of a defect propagating along the worldsheet appropriate to their respective geometries.  Owing to the symmetry of the background and geometries used in those references, their results follow readily using the techniques laid out here.

In Eq.~\eqref{eqn:DFF}, the Nambu-Goto action cancels out of the normalized two-point function, because
\begin{align}
  \dblavg{F_{\mu\nu}(x)F_{\rho\sigma}(y)}_\calC &= \exp(-S_{\rm NG}[X(\calC)] - S_w^{(0)}(x,y))
  \intertext{and}
  \dblavg{1}_\calC &= \exp(-S_{\rm NG}[X(\calC)])\,.
\end{align}
Nevertheless the Nambu-Goto action still determines the geometry of the minimal action configuration.  The backreaction of the defect on the string geometry would modify the string's geometry too.  For the calculations in this paper, this effect is small because the geometry of the string on its own and the geometry of the particle on its own can be computed and seen to lie close to each other.

Our second change compared to the earlier AdS/CFT computations is to warp the background metric in order to impart confining behavior to the boundary gauge theory.  We use the asymptotically Euclidean $\AdS_{d+1}$ ``metric wall'' background of Ref.~\cite{Andreev:2006ct}:
\begin{equation}\label{eqn:ConfiningBackground}
  ds^2 = g_{mn} dx^m dx^n = e^{4\Lambda^2 z^2}\, \frac{R^2}{z^2} \,(dz^2 + \delta_{\mu\nu}\,dx^\mu dx^\nu)\,,
\end{equation}
and orthonormal frame fields $v_m{}^a = (R/z)\,e^{2\Lambda^2z^2}\delta_m{}^a$. In Ref.~\cite{White:2007tu}, this background is compared with alternative metrics in its ability to compute rectangular Wilson loops, and is shown to provide the best agreement with lattice computations of the same.  As in Refs.~\cite{Andreev:2006ct} and \cite{White:2007tu}, we determine our parameters by reproducing lattice results for the Cornell potential between a heavy quark-antiquark pair modeled as a rectangular line current.  Doing so yields $\Lambda \approx 330\,{\rm MeV}$ and the dimensionless string tension $\tau = R^2/2\pi\alpha^\prime \approx 0.1836$. 

\begin{figure}[t]
  \centerline{
    \includegraphics[width=9cm]{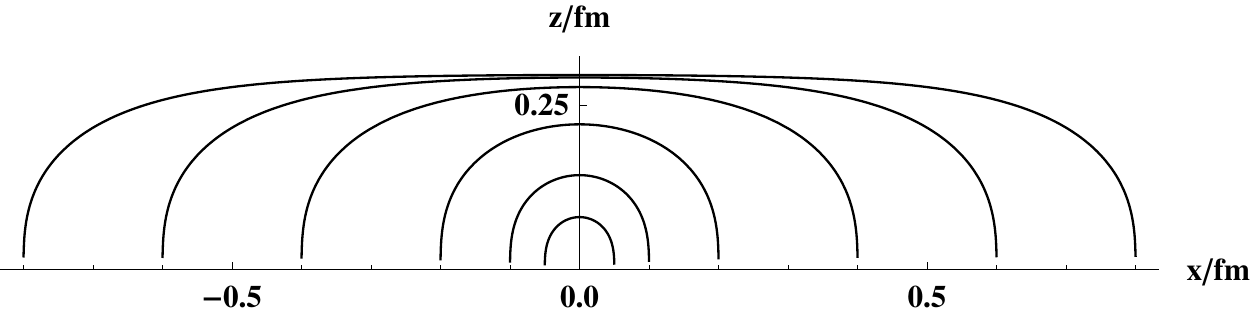}
  }
  \caption{String profiles with various values of $r$}
  \label{fig:ProfileSaturation}
\end{figure}

To decide which Wilson loop to use in our calculations, we consider our benchmark, the lattice computations of Ref.~\cite{DiGiacomo:1992df,DiGiacomo:1996mn} in which the field strength correlator is actually computed as a weighted sum over the results obtained with different Wilson loop contours.  So in fact there is no single contour $\calC$ which should exactly reproduce the lattice data.  However, the behavior of a particle propagating within a worldsheet will be nearly identical for all choices with geometries both much shorter than and much larger than the confinement scale in extent.  Slight differences would be allowed in the intermediate region but will not significantly affect the calculation.  So we choose the contour for convenience of calculation.  Two possible contours are shown in Fig.~\ref{fig:Contours}, and we use the one on the right.  We also take $x$ and $y$ to be separated only in the time direction $t$ by a boundary distance of $|x^\mu-y^\mu| = r$.  The profile is translationally invariant in some boundary direction perpendicular to $t$.  When parametrized by the time coordinate $t$, the radial coordinate $Z_0^z(t)$ of the minimum action path satisfies \cite{Andreev:2006ct}
\begin{equation}\label{eqn:StringProfile}
  \frac{e^{4\Lambda^2 Z_0^z(t)^2}}{Z_0^z(t)^2\sqrt{1 + \dot Z_0^z(t)^2}} = \frac{e^{4\Lambda^2 z_m^2}}{z_m^2}\,,
\end{equation}
where $z_m$ is the maximum value the string's profile assumes, and is determined implicitly by
\begin{equation}
  r/2 = \int_0^{z_m}\!dz\,\bigg(\frac{z_m^4}{z^4}\,e^{8\Lambda^2(z_m^2-z^2)} - 1\bigg)^{-1/2}\,.
\end{equation}
For $r \gg \Lambda^{-1}$, $z_m \approx z_\Lambda = (2\Lambda)^{-1}$.  That is, over long distances the worldsheet sits at the radial coordinate which minimizes $e^{4\Lambda^2z^2}/z^2$, which we call $z_\Lambda$.  An increase in $r$ serves only to lengthen this portion of the worldsheet, giving rise to linear behavior in the boundary theory's heavy quark potential.  Such behavior is common to all choices of contour $\calC$, with differences lying only in minor details of the exact profile.  For $r \ll \Lambda^{-1}$, a change in $r$ results in a new profile which is just a rescaled version of the old profile, due to asymptotic conformality near the boundary $z = 0$.  The shape of this profile is shown for several values of $r$ in Fig.~\ref{fig:ProfileSaturation}.
The particle's minimum action now has a position-dependent coupling to the curvature:
\begin{equation}\label{eqn:ClassicalActionNonconformal}
  S_w^{(0)}(x,y\,|\,s) = \frac{\sigma(x,y)^2}{2s} + \frac{s}{2}\int_0^1(m^2 - \kappa_{d,p}\calR(Z_0(\lambda)))\,d\lambda\,.
\end{equation}
The field strength operator is a 2-form, so for $d=4$ there is a simplification because $\kappa_{4,2} = 0$.  Moreover, its bulk mass $m$ vanishes.  This can be determined in the near boundary region, where the standard AdS/CFT scaling dimension formula Eq.~\eqref{eqn:ScalingDimFormula} holds.  In this regime, $\Delta = 2$ and so we find $m=0$.  As $m$ is not a function of position, it therefore vanishes everywhere.  But even if we set $m$ to zero, a term just like it will be inserted as an infrared regulator in Eq.~\eqref{eqn:SemiclassicalGreenFunc}.  The regulator must be kept until after we take the large $\sigma(x,y)$ limit of Eq.~\eqref{eqn:ClassicalActionNonconformal}, after which it is absorbed into the normalization.  After taking these steps, we find
\begin{equation}
  \exp(-S_w^{(0)}[Z_0]) \rightarrow \widetilde\calN\,\sigma(x,y)^{-d/2}\,.
\end{equation}
$D_{\rm VM}(x,y)^{1/2}$ was expressed using the determinant of the operator $\delta_{ab}\d_\lambda^2 + M_{ab}$ in Eq.~\eqref{eqn:AdSVMD}.
Unlike before, the curvature is no longer homogeneous and isotropic so that $M_{ab}$ is no longer $\delta_{ab}$ times a constant as it is in Eq.~\eqref{eqn:AdSTidalTensor}.  However, for $r \ll \Lambda^{-1}$ the path lies in the asymptotically conformal region and $M_{ab}$ will be well approximated by Eq.~\eqref{eqn:AdSTidalTensor}.  Furthermore, for $r \gg \Lambda^{-1}$ we will see that $M_{zz} \approx -2e^{-1}\sigma(x,y)^2/R^2$ and all other components will be small along the majority of the path.  The variation of $M_{ab}$ with $\lambda$ in the intermediate region is slow so we expand in $(\d_\lambda^2)^{-1}M_{ab}$ using the identity $\det(1 + \eepsilon) = 1 + \tr \eepsilon + O(\eepsilon^2)$, to write
\begin{align}
  \det^\prime(&-\delta_{ab}\d_\lambda^2 - M_{ab})\\
  &= (-\d_\lambda^2)^{d} - (-\d_\lambda^2)^{d-1}\, \tr M + O(M^2)\nonumber\\
  &= (-\d_\lambda^2 - d^{-1}\tr M)^{d} + O(M^2)\,.\nonumber
\end{align}
The rest of the calculation proceeds as before.  One could solve the differential equation of the Gel'fand-Yaglom theorem numerically but the slow variation of $\tr M$ allows us to write
\begin{align}\label{eqn:WKBVMdet}
  D_{\rm VM}&(x,y|s)^{1/2} \approx\\ &\frac{1}{(2\pi s)^{(d+1)/2}}\bigg(\frac{\sinh(\tint_0^1 \sqrt{-d^{-1}\tr M(\lambda)}\,d\lambda)}{\sqrt{-d^{-1}\tr M(0)}}\bigg)^{-d/2}\,.\nonumber
\end{align}
Following Eq.~\eqref{eqn:AffineParam}, define for the affine parametrization path $Z_0$ the normalized tangent $\hat t^a(\lambda) = \dot Z_0^a(\lambda)/\sigma(x,y)$. Then introduce for notational convenience,
\begin{equation}\label{eqn:theta}
  \cos\theta(\lambda) = \hat t^{\,x}(\lambda) \quad\text{and}\quad \sin\theta(\lambda) = \hat t^{\,z}(\lambda)\,.
\end{equation}
\begin{widetext}
  The Riemann tensor of the metric in Eq.~\eqref{eqn:ConfiningBackground} has components, in the orthonormal frame basis, given by
  \begin{equation}
  \begin{aligned}
    \calR_{abcd} &= -\frac{1}{R^2}\,e^{-4\Lambda^2z^2}\,(1-4\Lambda^2z^2)^2\,(\delta_{ac}\delta_{bd}-\delta_{ad}\delta_{bc})\,, \hspace*{1cm} && (a,b,c,d\neq z)\\
    \calR_{a z b z} &= -\frac{1}{R^2}\,e^{-4\Lambda^2z^2}\,(1+4\Lambda^2z^2)\,\delta_{ab}\,, && (a,b \neq z)\label{eqn:AdSQCDRiemannTensor}
  \end{aligned}
  \end{equation}
  with all others vanishing except those related by symmetries of the Riemann tensor to the components above.  We find
  \begin{align}\label{eqn:sqrttrM}
    &\sqrt{-d^{-1}\tr M(Z_0(\lambda))}\; = \\&\hspace*{1cm}\frac{\sigma(x,y)}{R}\,e^{-2\Lambda^2Z_0^z(\lambda)^2}\sqrt{(1-d^{-1})[\cos^2\theta(\lambda)(1-4\Lambda^2Z_0^z(\lambda)^2)^2+\sin^2\theta(\lambda)(1+4\Lambda^2Z_0^z(\lambda)^2)] + d^{-1}(1+4\Lambda^2Z_0^z(\lambda)^2)}\nonumber
  \end{align}
\end{widetext}
As $x$ and $y$ approach the boundary, the Green function in Eq.~\eqref{eqn:WKBVMdet} assumes its asymptotic form
\begin{equation}\label{eqn:RegGreenFunc}
  G^{(FF)}_{\rm reg}(x,y|\calC) = \widetilde\calN \exp(-\Delta\cdot\sigma_{\rm reg}(x,y) / R)\,,
\end{equation}
where $\Delta$ is defined by
\begin{equation}\label{eqn:Delta}
  \Delta(Z_0(\lambda)) = R\,\frac{d}{2}\sqrt{-d^{-1}\,\tr M(Z_0(\lambda))}
\end{equation}
and
\begin{equation}\label{eqn:sigmadotDelta}
  \sigma_{\rm reg} \cdot \Delta/R = \frac{\sigma_{\rm reg}(x,y)}{R}\int_0^1 \Delta(Z_0(\lambda))\,d\lambda\,.
\end{equation}
Using an affine parameter in the expressions above allowed the factorization of the distance $\sigma(x,y)$ out of the integral in Eq.~\eqref{eqn:sigmadotDelta}.  For a generic parameter $\xi$ of $Z_0$,
\begin{equation}
  \sigma_{\rm reg} \cdot \Delta/R = \frac{1}{R}\int_{\xi_x+\eepsilon}^{\xi_y-\eepsilon} \Delta(Z_0(\xi))\,\|\dot Z_0(\xi)\|\,d\xi\,,
\end{equation}
where $\|\dot Z_0(\xi)\| = \sqrt{\delta_{ab}\,\dot Z^a_0(\xi)\,\dot Z^b_0(\xi)}\,$, $Z_0(\xi_x) = x$ and $Z_0(\xi_y) = y$.  We will perform calculations using the $t$-coordinate parametrization of the contour shown on the right in Fig.~\ref{fig:Contours}, meaning we use $Z_0$ given in Eq.~\eqref{eqn:StringProfile}, for which
\begin{equation}\label{eqn:sigmaDotDeltatparam}
  \sigma_{\rm reg}\cdot\Delta/R = \!\!\int\limits_{-r/2+\eepsilon}^{r/2-\eepsilon}\!\!\Delta(Z_0(t))\,\frac{e^{2\Lambda^2 Z_0^z(t)^2}}{Z_0^z(t)}\,\sqrt{1+\dot Z_0^z(t)^2}\,dt\,
\end{equation}
and
\begin{align}
  \cos^2\theta(t) &= \frac{1}{1+\dot Z_0^z(t)^2}\\
  &= (Z_0^z(t)/z_m)^4\,\exp(8\Lambda^2z_m^2 - 8\Lambda^2Z_0^z(t)^2)\,.\nonumber
\end{align}

\begin{figure*}[t]
  \centering
  \centerline{
    \includegraphics[height=10cm]{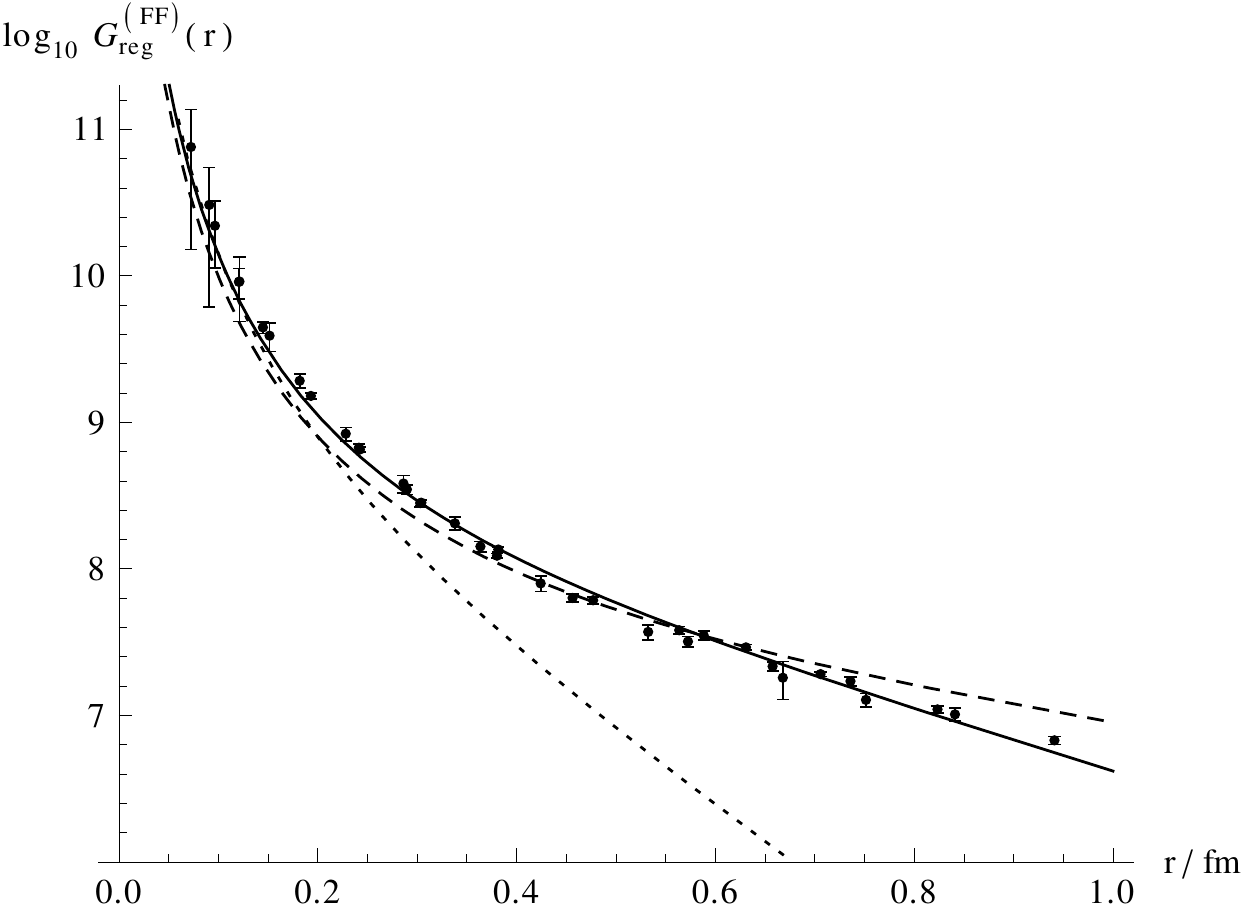}
  }
  {\tighten
    \caption[1]{Boundary-to-boundary field strength correlator with the correct $\Delta$ (solid) from Eq.~\eqref{eqn:Delta} contrasted with the incorrect $\Delta$ (dashed) of Eq.~\eqref{eqn:wrongDelta}, and $\Delta = 2$ held constant (dotted).  Lattice data was taken from Refs.~\cite{DiGiacomo:1992df,DiGiacomo:1996mn} and error bars from Ref.~\cite{Meggiolaro:2012}.}
    \label{fig:BndyBndyCorr}}
\end{figure*}

We can now compare the accuracy of the AdS/QCD calculations with lattice results.  For a 2-form, the index $\calA = [\mu\nu]$ is a pair of antisymmetrized vector indices.  By computing the term proportional to
\begin{equation}
  \delta_{[\mu\nu][\rho\sigma]} = \delta_{\mu\rho}\delta_{\nu\sigma} - \delta_{\mu\sigma}\delta_{\nu\rho}\,,
\end{equation}
we are comparing to the quantity called $D_\perp(x,y)$ in Ref.~\cite{DiGiacomo:1996mn}.  Our expression of this quantity comes from combining Eqs.~\eqref{eqn:sqrttrM}, \eqref{eqn:RegGreenFunc}, \eqref{eqn:Delta} and \eqref{eqn:sigmaDotDeltatparam} and setting $d = 4$.  Figure~\ref{fig:BndyBndyCorr} shows the result plotted against the lattice data \cite{DiGiacomo:1992df,DiGiacomo:1996mn, Meggiolaro:2012}.  Owing to the existence of a background, other tensor structures are possible and given in that reference \footnote{To compute them in our formalism, we would need to move beyond the picture of a free field in a warped background by coupling the bulk $2$-form field to other fields.  That calculation is outside the scope of the current paper.}.  The solid curves show the propagator Eq.~\eqref{eqn:RegGreenFunc}.  At small $r$, the asymptotically conformal behavior dictates $G(r) \propto 1/r^{2\Delta_{\rm UV}}$, where $\Delta_{\rm UV} = \Delta(z \to 0) = 2$ for the field strength operator.  The dotted curve in Fig.~\ref{fig:BndyBndyCorr} shows the propagator computed with the scaling dimension fixed everywhere at the ultraviolet value $\Delta_{\rm UV}$.  The normalizations of the dashed and dotted curves were set to agree with the lattice data in the small $r$ limit, where the universal $1/r^{2\Delta_{\rm UV}}$ behavior must hold.  For the solid curve, the normalization is determined by minimizing the $\chi^2$ value of the fit.  The end result gives $\chi^2/{\rm d.o.f.} \approx 3.07$.

For large $r$, exponential behavior results from saturating the $z_m \leq z_\Lambda$ bound:
\begin{equation}
  G^{(FF)}(x,y) \propto \exp(-|x-y|/\lambda_{\rm gl})\,,
\end{equation}
where the gluonic correlation length is
\begin{equation}
  \lambda_{\rm gl} = R\,(\Delta(z_\Lambda)\sqrt{g_{00}(z_\Lambda)}\,)^{-1} = (2\sqrt{2}\,\Lambda)^{-1}\,.
\end{equation}
Substituting $\Lambda \approx 330\,{\rm MeV}$, we find that
\begin{equation}
  \lambda_{\rm gl} \approx 0.21\,{\rm fm}\,.
\end{equation}
Our result is in good agreement with the lattice results of Ref.~\cite{DiGiacomo:1992df,DiGiacomo:1996mn}, which indicate $\lambda_{\rm gl} \approx 0.22\,{\rm fm}$.  Reference~\cite{Bali:1997aj} has $\lambda \approx 0.11\text{-}0.13\,{\rm fm}$, different roughly by a factor of $2$.  But one of the same authors later computes the effective mass of the field strength propagator at long distances to be around 887 MeV \cite{Bali:2003jq}.  This quantity is the reciprocal of the the correlation length, implying that $\lambda_{\rm gl} \approx 0.225\,{\rm fm}$.  Reference~\cite{Juge:2002br} computes excited Wilson loop potentials and indicates a similar value for the effective mass of the lightest excitation in its data.

As a final note, we address a technical point.  When using a generalized proper time regularization of multidimensional path integrals, one must keep in mind that the regularization of functional determinants does not commute with taking the determinant over finite indices \cite{Visser:1992pz,Evans:1998pd,Filippi:1998fe}.  We have ordered the functional and algebraic determinants as shown in Eq.~\eqref{eqn:DetOfAPower}.  Some authors \cite{Visser:1992pz,Dowker:1998tb} claim either explicitly or implicitly that it is correct to take the functional determinant of $\delta_{ab}\d_\lambda^2 + M_{ab}$ before taking the determinant over finite indices, i.e.\ computing
\begin{equation}\label{eqn:detDet}
  \bigg(\frac{\det^\prime\Det(-\delta_{ab}\d_\lambda^2 - M_{ab})}{\det^\prime\Det(-\delta_{ab}\d_\lambda^2)}\bigg)^{-1/2}\,.
\end{equation}
To evaluate Eq.~\eqref{eqn:detDet}, one can use a multidimensional Gel'fand-Yaglom theorem \cite{Visser:1992pz},
\begin{equation}
  \frac{\det^\prime\Det(-\delta_{ab}\d_\lambda^2 - M_{ab})}{\det^\prime\Det(-\delta_{ab}\d_\lambda^2)} = \det A(1)^{-1}
\end{equation}
where $A^a{}_b(\lambda)$ solves
\begin{equation}
  (\delta_{ab}\,\d_\lambda^2 + M_{ab})A^b{}_c(\lambda) = 0
\end{equation}
with boundary conditions $A^a{}_b(0) = 0$ and $\dot A^a{}_b(0) = \delta^a{}_b$.  We will not pursue the details, but using a WKB approximation of this differential equation results in a propagator of the form in Eq.~\eqref{eqn:RegGreenFunc} except with
\begin{equation}\label{eqn:wrongDelta}
  \Delta = \tfrac{1}{2}R\,\tr\sqrt{-M(Z_0)}
\end{equation}
instead of Eq.~\eqref{eqn:Delta}.  Reference~\cite{Filippi:1998fe} shows that the order used in the present paper is the correct order.  Furthermore, we see Eq.~\eqref{eqn:wrongDelta} does not give a propagator in good agreement with the lattice data.  Its use results in the dashed curve of Fig.~\ref{fig:BndyBndyCorr}.

  In this paper, we have addressed two closely related problems.  We set out to compute the two-point function of the gluon field strength operator known from quenched lattice QCD computation.  To perform this calculation, we first had to gain an understanding of the origin of the scaling dimension formula known from AdS/CFT.  From there, we learn how to extend the scaling dimension formula of AdS/CFT to warped backgrounds.  By restricting our attention to only one correlator, namely the field strength two-point function, with end points taken not to coincide, we avoid the need for contact terms.
  
The results presented here should open up a number of lines of further inquiry, including how to treat higher-order correlation functions or correlation functions of more complicated operators, or relaxing the approximations contained in this paper.

\begin{acknowledgements}
  This work was supported in part by the Director, Office of Science, Office of Nuclear Physics, of the U.S. Department of Energy under Grant No.\ DE-FG02-05ER41368. The author thanks the Department of Energy's Institute for Nuclear Theory at the University of Washington for its hospitality during the early stages of this work.  The author thanks O.\ Andreev, E.\ Meggiolaro, C.\ Morningstar, B.\ Mueller, R.\ Plesser and V.\ Rychkov for fruitful discussion and correspondence, as well as the anonymous journal referee for many helpful comments which helped refine the document, and also would like to thank his advisor T.\ Mehen in particular for asking hard questions.
\end{acknowledgements}


\bibliographystyle{unsrt}
\bibliography{final}{}
\end{document}